\documentclass[useAMS,usenatbib]{mn2e}
\usepackage{graphicx}
\usepackage{natbib, aas_macros}
 
\title[Source counts from the AKARI NEP survey]{The 2 to 24 micron source counts from the AKARI North Ecliptic Pole survey}
\author[K.Murata et al.]{
  K.Murata$^{1,2}$
  \thanks{E-mail:murata@ir.isas.jaxa.jp},
  C.P.Pearson$^{3,4,5}$,
  T.Goto$^{6}$,
  S.J.Kim$^{7}$,
  H.Matsuhara$^{1,2}$,
  and T.Wada$^{1}$
  \\
  $^{1}$Institute of Space and Astronautical Science, Japan Aerospace Exploration Agency, Sagamihara, 229-8510 Kanagawa, Japan\\
  $^{2}$Department of Space and Astronautical Science, The Graduate University for Advanced Studies, Japan\\
  $^{3}$ Astrophysics Group, Department of Physics, The Open University, Milton Keynes, MK7 6AA, UK\\
  $^{4}$ RAL Space, Rutherford Appleton Laboratory, Chilton, Didcot, Oxfordshire OX11 0QX, UK\\
  $^{5}$ Oxford Astrophysics, Oxford University, Keble Road, Oxford OX1 3RH, UK\\
  $^{6}$ Institute of Astronomy and Department of Physics,National Tsing Hua University,\\
  No. 101, Section 2, Kuang-Fu Road, Hsinchu 30013, Taiwan, R.O.C\\
  $^{7}$ Department of Physics and Astronomy, Seoul National University, Shillim-Dong Kwanak-Gu, Seoul 151-742, South Korea
}
\begin{document}

\date{Accepted 2014 August 07. Received 2014 July 28; in original form 2014 May 26}

\pagerange{\pageref{firstpage}--\pageref{lastpage}} \pubyear{2014}

\maketitle

\label{firstpage}

\begin{abstract}
  We present herein galaxy number counts of the nine bands in the 2--24 micron range on the basis of the {\it AKARI} North Ecliptic Pole (NEP) surveys.
  The number counts are derived from NEP-deep and NEP-wide surveys, which cover areas of 0.5 and 5.8 deg$^2$, respectively.
  To produce reliable number counts, the sources were extracted from recently updated images.
  Completeness and difference between observed and intrinsic magnitudes were corrected by Monte Carlo simulation.
  Stellar counts were subtracted by using the stellar fraction estimated from optical data.
  The resultant source counts are given down to the 80\% completeness limit; 0.18, 0.16, 0.10, 0.05, 0.06, 0.10, 0.15, 0.16, and 0.44 mJy in the 2.4, 3.2, 4.1, 7, 9, 11, 15, 18 and 24 $\mu$m bands, respectively.
  On the bright side of all bands, the count distribution is flat, consistent with the Euclidean Universe, while on the faint side, the counts deviate, suggesting that the galaxy population of the distant universe is evolving.
  These results are generally consistent with previous galaxy counts in similar wavebands.
  We also compare our counts with evolutionary models and find them in good agreements.
  By integrating the models down to the 80\% completeness limits, we calculate that the {\it AKARI} NEP-survey revolves 20\%-50\% of the cosmic infrared background, depending on the wavebands.
  
\end{abstract}

\section{Introduction}
To understand galaxy evolution and the history of cosmic star formation, galaxy number counts are typically investigated as a function of apparent magnitudes \cite[]{1999A&A...351L..37E,2004ApJS..154...80C,2004ApJS..154...70P,2010A&A...514A...8P,2012MNRAS.423..575S}.
In particular, infrared source counts are quite important because most of the energy from star formation is absorbed by dust and re-radiated in the infrared.
The 15 $\mu$m source counts provided by the Infrared Space Observatory (ISO) revealed a strong evolution in the redshift $z$$<$1.5 \cite[]{1999A&A...351L..37E}.
This evolution was confirmed by the Spitzer 24 $\mu$m source counts \cite[]{2004ApJS..154..112L,2004ApJS..154...70P}, which implies that the infrared output is dominated at $z$=0.5--2.5.
In addition, evolutionary models have also been produced to fit source counts \cite[]{1996MNRAS.283..174P,2005MNRAS.358.1417P,2009MNRAS.394..117R,2011A&A...529A...4B,2013ApJ...768...21C}.
The models in \cite{2005MNRAS.358.1417P} which reproduce the ISO and Spitzer counts indicate that the source counts depend strongly on the many mid-infrared dust features at 3.3, 6.2, 7.7, 8.6, 9.7, 11.3 and 12.7 $\mu$m.
Although these dust features are complicated to model, they are sensitive to the source counts \cite[]{2004ApJS..154..112L}.
Therefore, source counts with additional bands are invaluable to constrain the evolution of the dusty galaxy population.
\par
Thanks to the unique continuous wavelength coverage at 2--24 $\mu$m with nine photometric bands, the Japanese-led {\it AKARI} satellite \cite[]{2007PASJ...59S.369M} is capable of contributing to this field.
With all nine bands, {\it AKARI} made deep and wide surveys in the direction of the North Ecliptic Pole (NEP survey; Matsuhara et al. 2006, Wada et al. 2008, Takagi et al. 2012).
The survey images were recently revised to detect fainter objects and to be more reliable \cite[]{2012A&A...548A..29K,2013A&A...559A.132M}.
\par
In this study, we present galaxy number counts from the nine {\it AKARI} bands based on the revised NEP-survey data.
This paper is organised as follows:
Section 2 presents the data and its analysis.
In section 3, the source counts from the nine bands are presented and compared with results from previous studies and with evolutionary models.
Finally, this work is summarised in section 4.

\section{Data and Methods}
\subsection{AKARI NEP survey}
The {\it AKARI} NEP survey exploited the following nine photometric bands covered by Infrared Camera \cite[]{2007PASJ...59S.401O}: N2, N3, N4, S7, S9W, S11, L15, L18W and L24.
The labels 'N', 'S' and 'L' indicate the near-infrared (NIR), mid-infrared short (MIR-S), and mid-infrared long (MIR-L) channels of the IRC, respectively; the numbers give the reference wavelength, and 'W' indicates a wide band width.
The survey consists of the two surveys; NEP-deep \cite[]{2008PASJ...60S.517W} and NEP-wide \cite[]{2009PASJ...61..375L}.
The NEP-deep survey covers approximately 0.5 deg$^2$ of the sky and the NEP-wide survey covers 5.8 deg$^2$ surrounding the NEP-deep field.
The observations were conducted between April 2006 and August 2007.
\par
The NEP-deep survey images were recently revised by \cite{2013A&A...559A.132M}.
The previous images suffered from significant contamination due to the behaviour of the detector and the optical system.
In the revised images, scattered light and stray light from the Earth limb were removed and artificial patterns were appropriately corrected.
As a result, the five sigma detection limits have been improved to 0.011, 0.009, 0.01, 0.03, 0.03, 0.06, 0.09, 0.09 and 0.26 mJy in the N2, N3, N4, S7, S9W, S11, L15, L18W and L24 bands, respectively.
By using the NEP-deep images we can constrain the faint end of the source counts.
In addition, sources can be extracted with remarkably better reliability, which is crucial for the source-count study.
\par
The NEP-wide survey images were updated by \cite{2012A&A...548A..29K}, who removed the artificial sources created by bright objects. 
The large areal coverage of these images allows us to detect many bright objects so that they can be used to constrain the bright side of the source counts.
Although significant contamination remains in the NEP-wide images, the resulting false detection is negligible at fluxes $>$1 mJy (see section \ref{rely}).

\subsection{Source extraction and photometry}
Source extraction and photometry were done with SExtractor \cite[]{1996A&AS..117..393B}.
The detection threshold was carefully optimised to reduce the false detection ratio.
In the image map, we applied the criterion of ten (five) connected pixels with a signal over 3.5$\sigma$ (2.5$\sigma$) above the local background for the NEP-deep (NEP-wide) and used the noise and weight maps produced from the mosaicked images. 
We discarded the five arcmin radius region centred at (R.A. = 269.60858, Dec. = 66.62320) to avoid light scattered from NGC6543 (Cat$'$s Eye nebula), which caused a number of fake sources.
\par
\begin{figure*}
\centering
\begin{minipage}{0.38\hsize}
\includegraphics[]{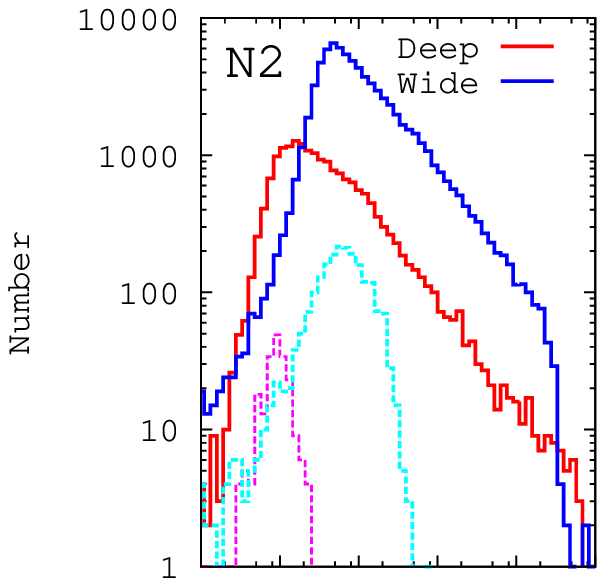}
\end{minipage}
\begin{minipage}{0.30\hsize}
\includegraphics[]{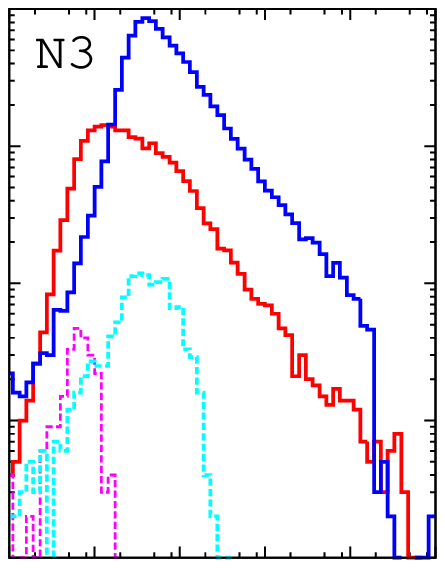}
\end{minipage}
\begin{minipage}{0.30\hsize}
\includegraphics[]{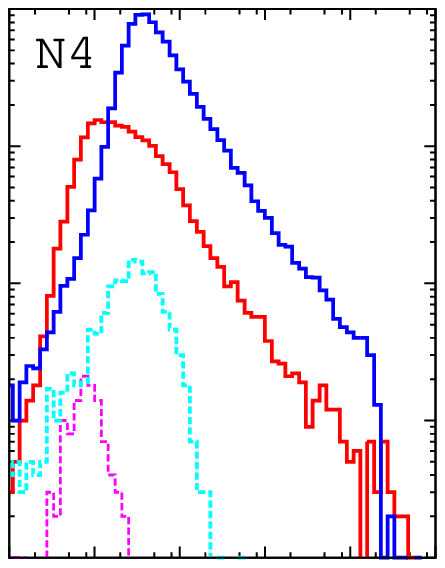}
\end{minipage}

\begin{minipage}{0.38\hsize}
\includegraphics[]{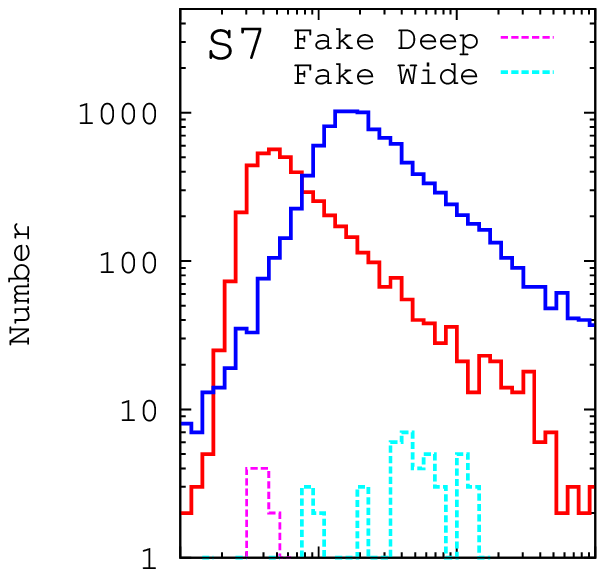}
\end{minipage}
\begin{minipage}{0.30\hsize}
\includegraphics[]{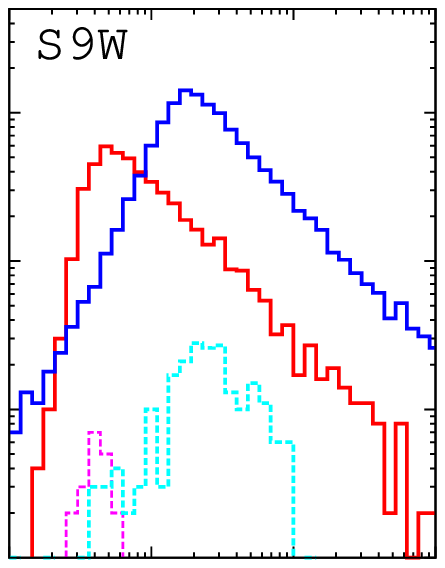}
\end{minipage}
\begin{minipage}{0.30\hsize}
\includegraphics[]{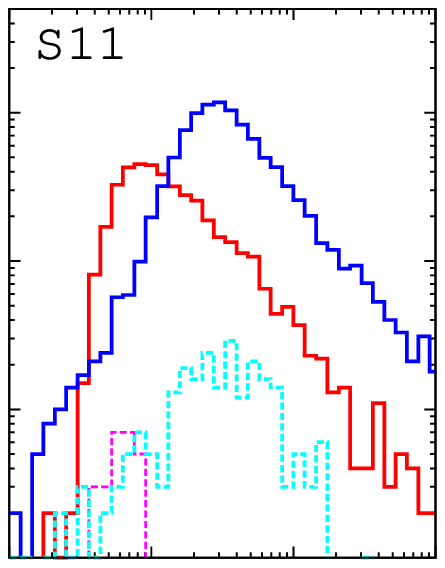}
\end{minipage}

\begin{minipage}{0.38\hsize}
\includegraphics[]{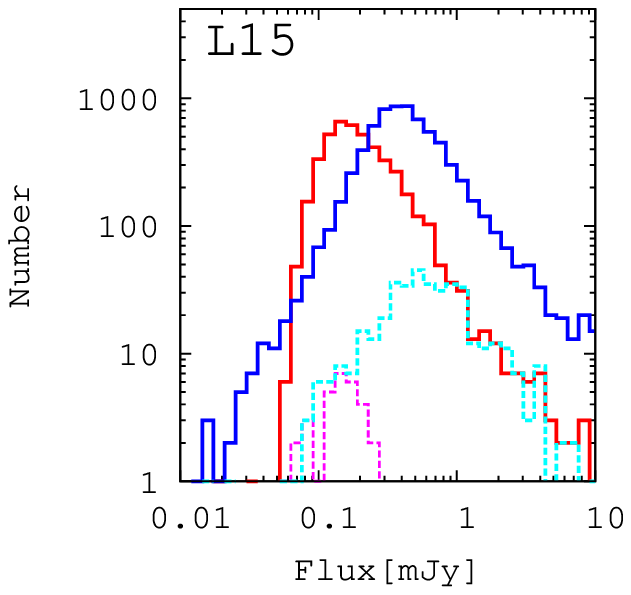}
\end{minipage}
\begin{minipage}{0.30\hsize}
\includegraphics[]{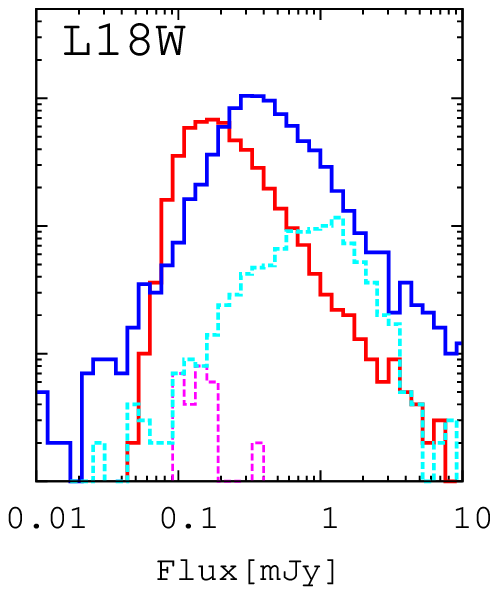}
\end{minipage}
\begin{minipage}{0.30\hsize}
\includegraphics[]{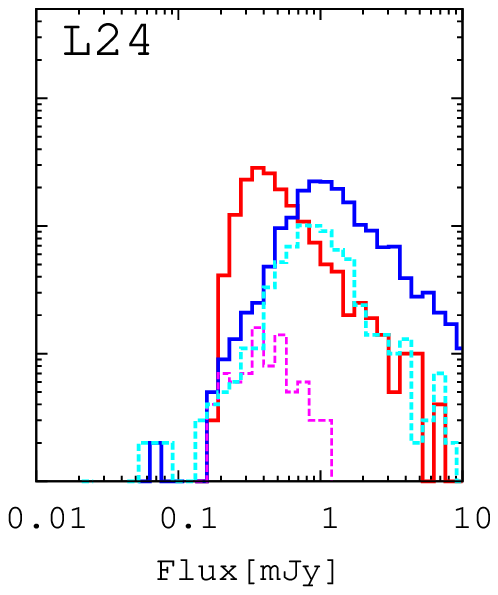}
\end{minipage}

\caption{Flux histogram of both NEP-deep and NEP-wide surveys for each band.
The red and blue lines indicate the number of objects detected in the NEP-deep and NEP-wide surveys, respectively.
The magenta and cyan lines indicate the fake sources from the negative images from the NEP-deep and NEP-wide surveys, respectively.
\label{fig:hist}}
\end{figure*}

For aperture photometry, the aperture radii were 6.3 arcsec for NIR and 6.0 arcsec for MIR-S and MIR-L of the NEP-deep survey, and 7.5 pixels for all bands of the NEP-wide survey.
The flux through the aperture was measured by subdividing the pixels into 5$\times$5 sub-pixels to measure all the flux passing through the aperture.
No aperture correction was needed because the flux calibrations were done by \cite{2013A&A...559A.132M} and \cite{2008PASJ...60S.375T} for these aperture radii, where the constant $\nu f_\nu$ was assumed to estimate the flux at the reference wavelength.
Fig. \ref{fig:hist} shows a histogram of the source flux extracted from all bands.
The results show that the NEP-deep survey (red) is two to four times deeper than the NEP-wide survey (blue), and owing to the large area, there are approximately ten times more sources in the NEP-wide survey. 
Therefore, the NEP-deep survey is useful to cover the fainter side of the source counts and the NEP-wide survey is better for the brighter side.
\par
Notably, although the source catalogue of the NEP-deep survey was published in \cite{2013A&A...559A.132M}, we extracted the sources again from the same images because the published catalogue is based on a detection image made by stacking all six MIR bands, which makes the detection limit deeper but makes the completeness depend on colour.
To avoid this effect, we extracted sources from single-band images.

\subsection{Reliability}
\label{rely}
The false detection rates were estimated on the basis of the same source extraction from the negative images.
If we assume that the false detections are due only to the random sky background deviation, then the number of the false detections should be the same in both the positive and negative images.
The flux histogram of such spurious sources (see Fig. \ref{fig:hist}) shows that sources extracted from NEP-deep images are quite reliable: sources in the N2-L18W and L24 bands are 99\% and 95\% reliable.
\par
\begin{figure*}
\centering
\includegraphics[]{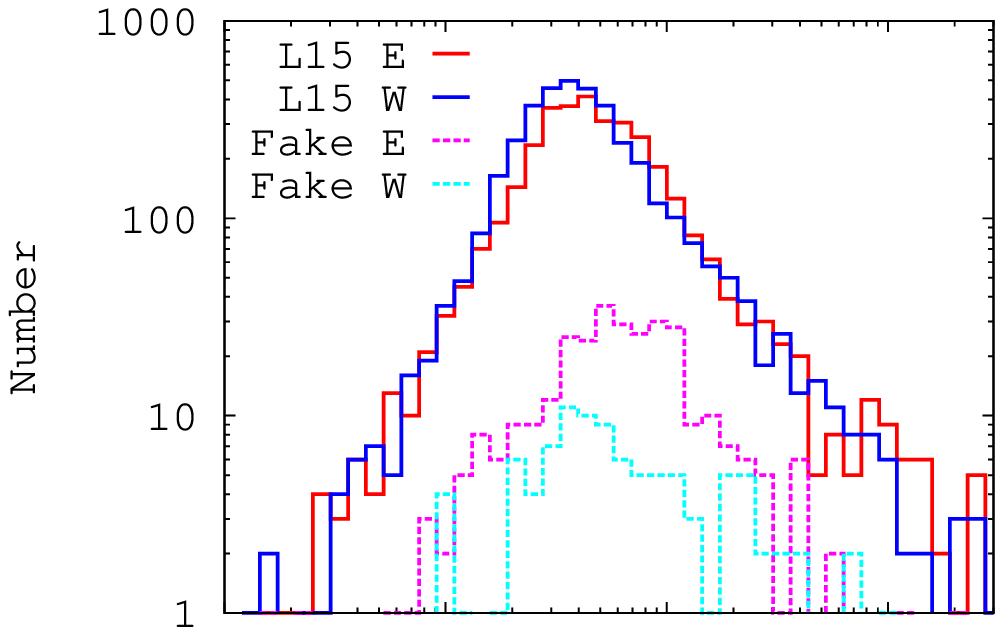}\\
\includegraphics[]{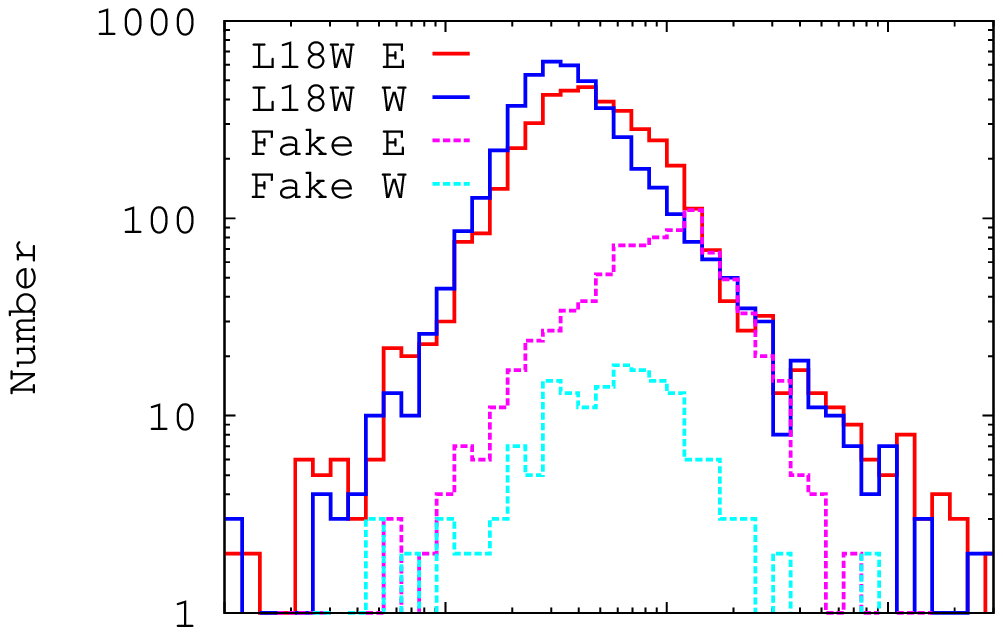}\\
\includegraphics[]{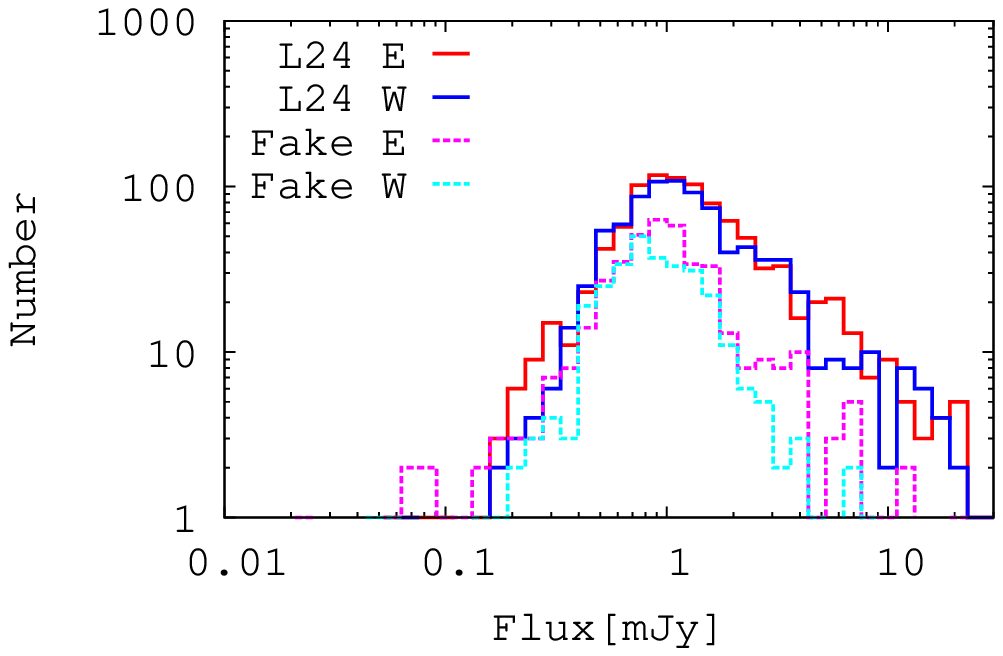}
\caption{
Number of sources from east and west sides of NEP-wide images.
The red and blue lines show the number of objects from the east and west side of positive images, respectively.
The magenta and cyan lines show the number of objects from the east and west side of negative images, respectively.
Although the fake number differs in both images, the positive numbers are the same on both sides of the images, which indicates the fake sources from the east side do not reflect the real false detection from the positive images.
\label{fig:wew}}
\end{figure*}

However, a significant number of spurious sources are detected in the NEP-wide images, especially in the MIR-L bands.
One of the main causes of false detection is stray light from the Earth limb, which is more intense at longer wavelengths.
In particular, the north-east parts of the images are contaminated because the north-east area was observed during the season when strong stray light from the Earth limb is incident on the telescope. 
Because light from the Earth limb causes systematic noise in addition to random noise, false detection in the negative images does not necessarily reflect fake sources in the positive images.
In this case, the false detection rate on the east side would be overestimated, leading to more fake sources on the east side of the negative images whereas the same number of sources should be found on both sides.
To check this possibility, we divided the NEP-wide images into east and west sides with the border at R.A. = 269.97.
The results are shown in Fig.\ref{fig:wew}.
As expected, more fake sources appear on the east side of the images than on the west side, and both images contain the same number of the positive sources, which indicates that the fake sources on the east side of the negative images do not reflect real false detection from the positive images.
Although many fake sources remain in the west side images, most of them are fainter than 1 mJy for the L15 and L18W bands.
Thus, we use objects only with $>$1 mJy from NEP-wide images for L15 and L18W bands.
In the present study, because the NEP-wide images are used only to constrain the brighter side of the source counts, this cut-off does not affect our results.
However, the numbers of spurious sources from both sides of the NEP-wide image at 24 $\mu$m are comparable, which we attribute to bad pixels.
  Because the sensitivity is lower at 24 $\mu$m, a significant number of pixels can be bad. 
Therefore, we do not use the L24 counts from the NEP-wide image because of its high rate of fake sources.

\subsection{Completeness}
To accurately estimate the source counts, two effects are required to be corrected: the lower detection rate of fainter objects and the difference between observed and intrinsic fluxes.
These effects were estimated through a Monte Carlo simulation with artificial sources for the NEP-deep.
Because the NEP-wide was used only for bright sources, we did not correct its completeness.
We injected artificial sources with the same radial profile as the point spread function into the original images and extracted them by using SExtractor with the same parameters.
To avoid self-confusion, the artificial sources were separated from each other by more than 60 pixels.
To reflect source confusion in the completeness, the real sources were not removed before the artificial sources were injected.
We regarded the sources to be successfully extracted when the position and the magnitude of the extracted sources agreed within 3.0 and 2.1 arcsec and 1 mag for the MIR and NIR bands, respectively.
The difference of 1 mag is large enough to avoid missing the source at the given magnitude.
The calculation was done at each magnitude with an interval of 0.2 mag.
This magnitude interval was selected to probe the completeness curve in detail.
For the MIR bands, we injected 100 sources per simulation and repeated each simulation 20 times.
However, because sources detected in the NIR band were limited by confusion, we injected only 20 sources per simulation in the NIR bands and repeated each simulation 100 times.
The completeness correction factors were defined as the ratio of the extracted number of sources to the injected number of sources at each flux bin.
\par
\begin{figure*}
\centering
\includegraphics[]{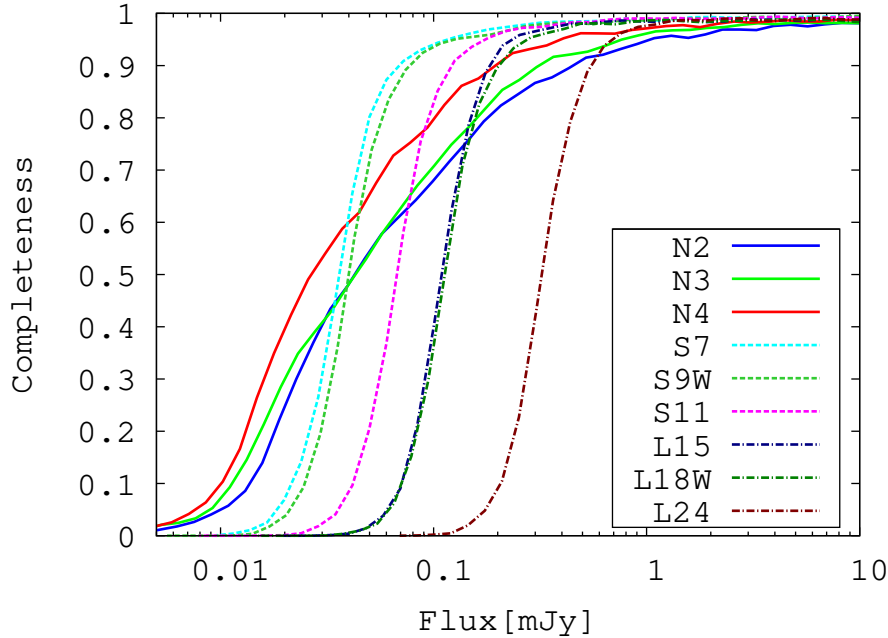}
\caption{Completeness of source extraction from each band of the NEP-deep images.
\label{fig:comp}}
\end{figure*}

The completeness curves for each band are shown in Fig. \ref{fig:comp}.
The 80\% completeness limits of the N2, N3, N4, S7, S9W, S11, L15, L18W and L24 bands are 0.18, 0.16, 0.10, 0.050, 0.058, 0.095, 0.15, 0.16 and 0.44 mJy, respectively.
In the next section, these limits are applied as the flux cuts of our source counts.
They are slightly shallower than those in the published catalogue \cite[]{2013A&A...559A.132M}, which is because we did not use detection images and the sources were extracted from each single band image to avoid complications of the completeness correction.
The completeness curves of the MIR-S and MIR-L bands are dropped rapidly at the 50\% completeness limit, while those of NIR bands are gradually decreased.
This is because the detection limits of the MIR-S and MIR-L bands were determined by sky noise and those of the NIR bands by the confusion.
\par
A $P_{ij}$ matrix was produced to correct the observed and intrinsic flux differences \cite[]{1995ApJ...449L.105S,2004ApJS..154...80C}.
The elements of the $P_{ij}$ matrix give the probability that an object with $m_i$ is observed with $m_j$.
The matrix was normalised for the sum over $i$ for a given $j$ to give the observed number $N_j$.
Then, the corrected source count $N_i$ in each flux bin $i$ is the sum over $j$ of the re-normalised matrix.
However, these details do not significantly affect our results because they are based on sources brighter than the 80\% completeness limits, where the signal-to-noise ratio is high so that few objects were observed with a significantly different flux.

\subsection{Stellar fraction}
\label{fste}
To determine the number of detected galaxies, the stellar fraction was estimated to be a function of the flux for each band by using optical data taken by Canada France Hawaii Telescope (CFHT), and the Guide Star Catalogue (GSC), version 2.3.2 \cite[]{2008AJ....136..735L}.
The CFHT data cover 2 deg$^2$ ($\sim$35\%) of the NEP-wide field \cite[]{2007ApJS..172..583H}, whereas the GSC data cover the entire field.
The stellar sources were identified on the basis of the criteria of stellarity $>$ 0.8 and $r'$ $<$ 19 \cite[]{2012A&A...548A..29K} from the CFHT catalogue and flag = 0 in GSC2.3.
The {\it AKARI} sources located in the area covered by CFHT were cross-matched with these star catalogues with a search radius of 2.1 and 3.0 arcsec for NIR and MIR-S and MIR-L bands, respectively, which is the same as Murata et al.(2013).
The stellar fraction was defined as the ratio of the number of matched {\it AKARI} sources to the number of {\it AKARI} sources located in the area covered by CFHT.
Notably, one may consider that the criterion $r'$ $<$ 19 underestimates the stellar fraction of fainter sources, but this has little effect on the MIR-S and MIR-L counts.
  However, this criterion could decrease the N2 counts at $S$$\sim$0.2 mJy by approximately 0.2 dex and the N3 and N4 counts at $S$$\sim$0.2mJy by approximately 0.1 dex.
  These reductions should be kept in mind as uncertainties in stellar subtraction.

\par
\begin{figure*}
\centering
\includegraphics[]{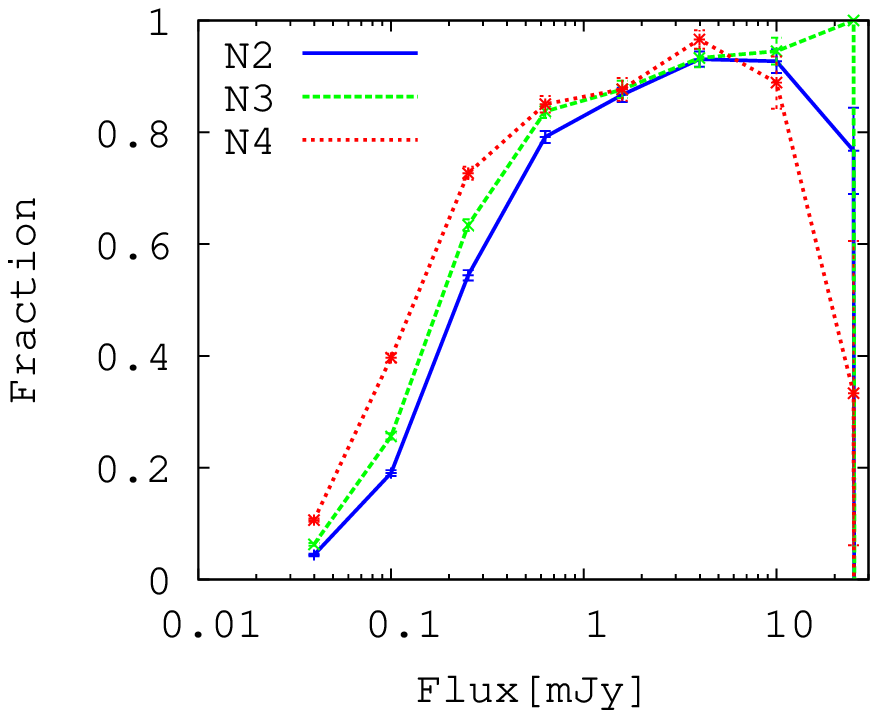}
\\
\includegraphics[]{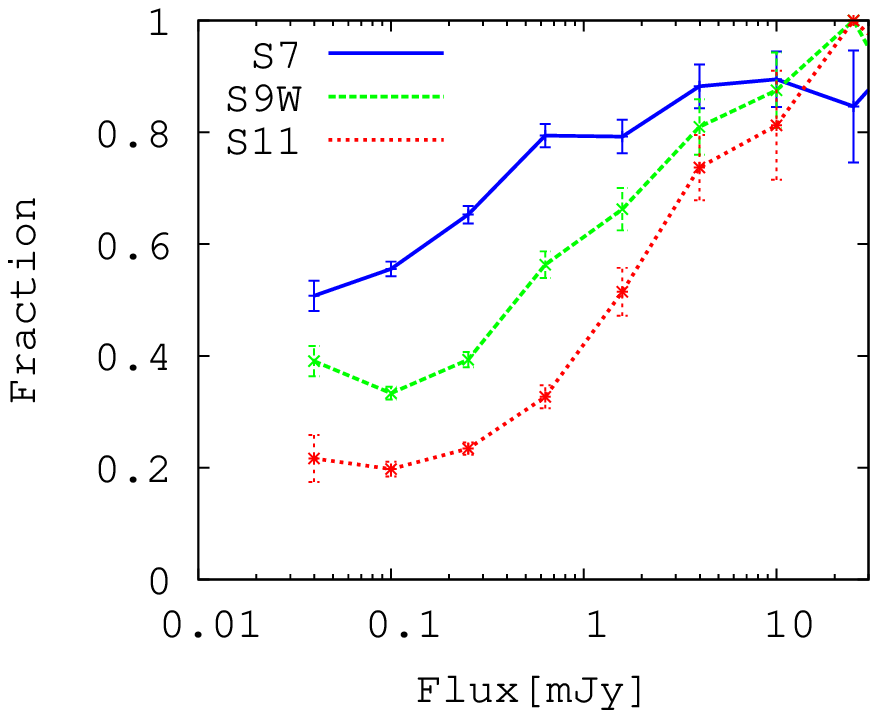}
\includegraphics[]{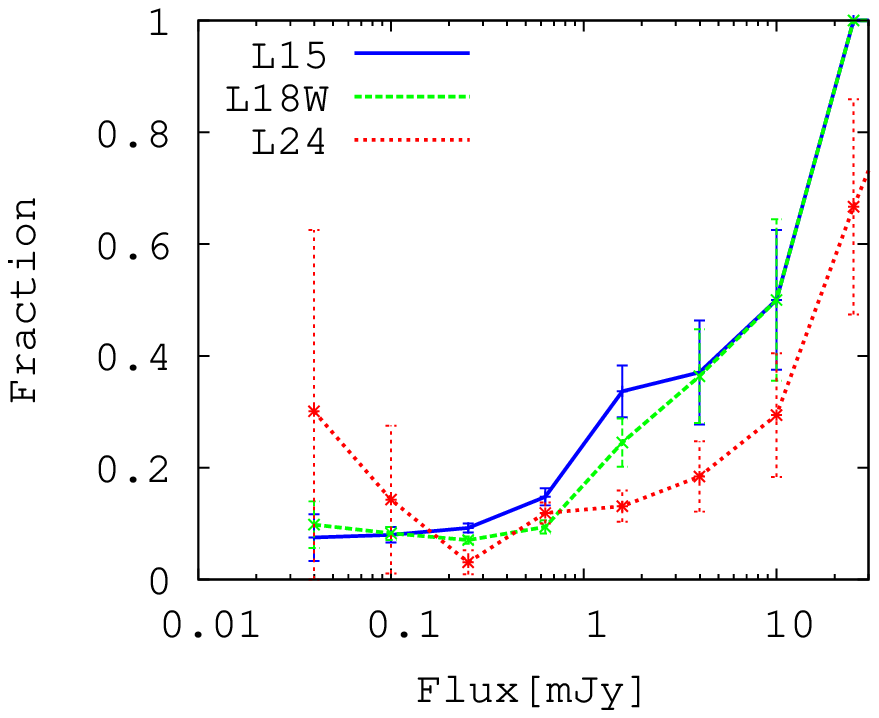}
\caption{Stellar fraction as functions of fluxes for each band.
  From top to bottom, stellar fraction of the NIR, MIR-S, and MIR-L bands are shown.
  The error bars were estimated with binomial distribution.
\label{fig:starfrac}}
\end{figure*}

Figure \ref{fig:starfrac} shows the stellar fraction as a function of flux for each {\it AKARI} band.
The error was calculated with the binomial distribution.
The results show that the fraction is higher for higher flux and shorter wavebands.
For the NIR bands, the stellar fraction is lower for the highest flux, which may be attributed to the small number statistics.
Notably, although the stars should ideally be removed individually from the catalogue, we had to use stellar fraction for the star subtraction on account of the areal coverage of the optical catalogue.
The stellar fraction at a given flux is estimated by linear interpolation.

\section{Results and Discussion}
\subsection{Source counts}
\begin{figure*}
\centering
\includegraphics[]{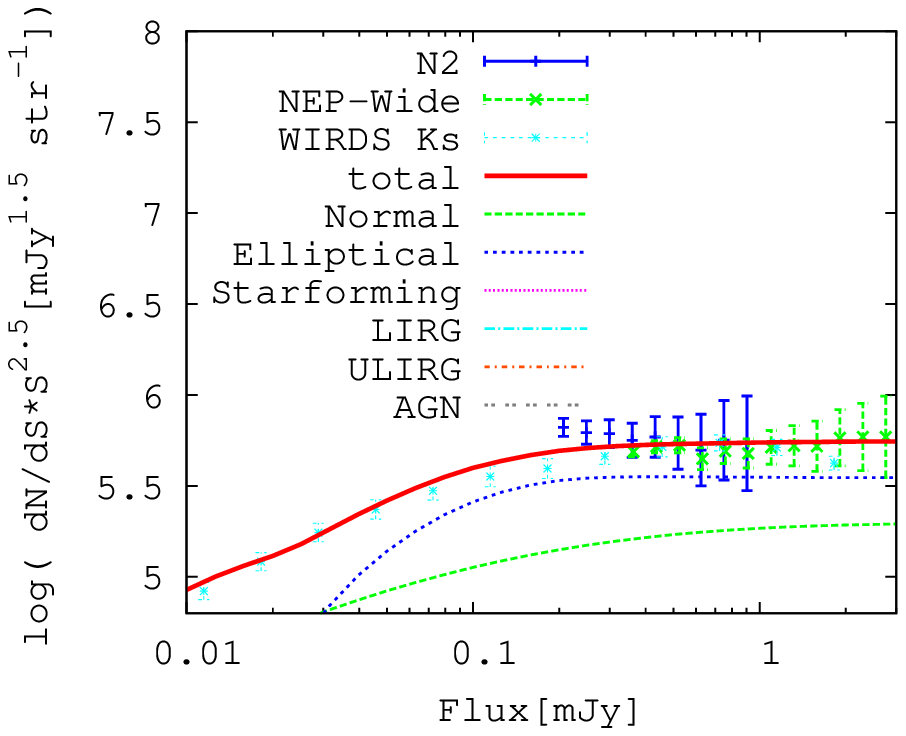}\\
\includegraphics[]{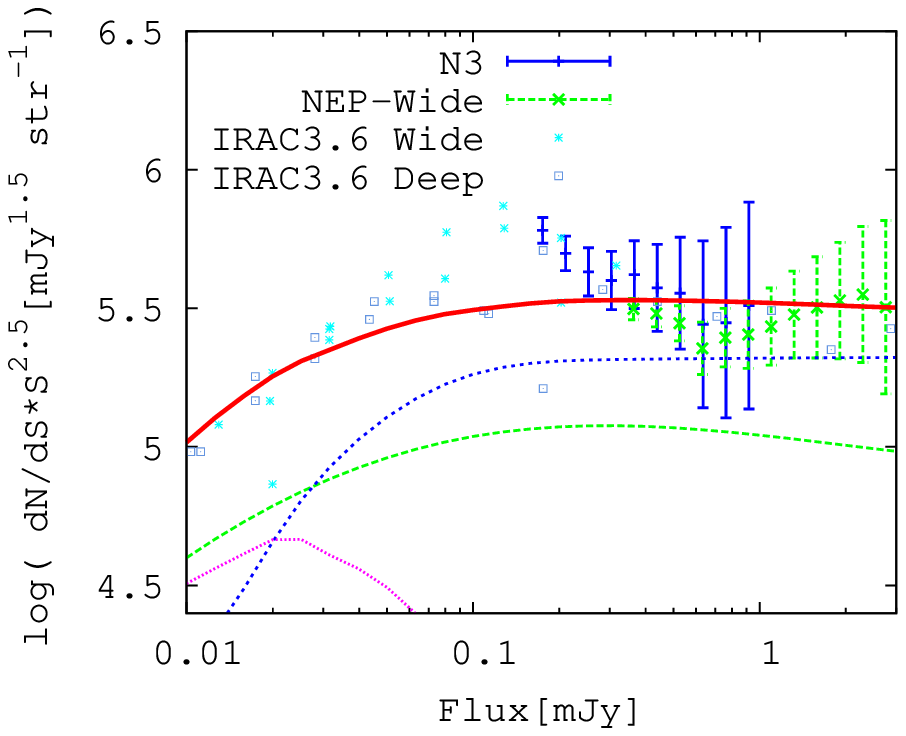}\\
\includegraphics[]{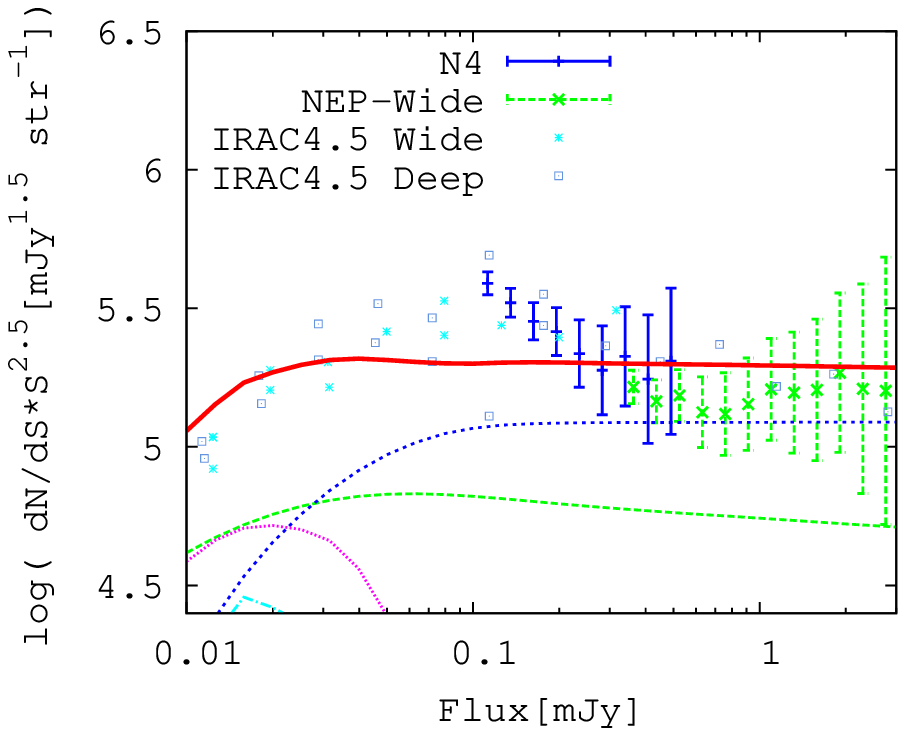}\\
\caption{Differential-source counts from all NIR bands.
  The evolutionary model is also shown with the components:
  Total, normal, elliptical, star-forming, LIRG, ULIRG and AGN are shown with red, green, blue, magenta, cyan, orange, and grey lines,respectively.
\label{fig:numcntn}
}
\end{figure*}

\begin{figure*}
\centering
\includegraphics[]{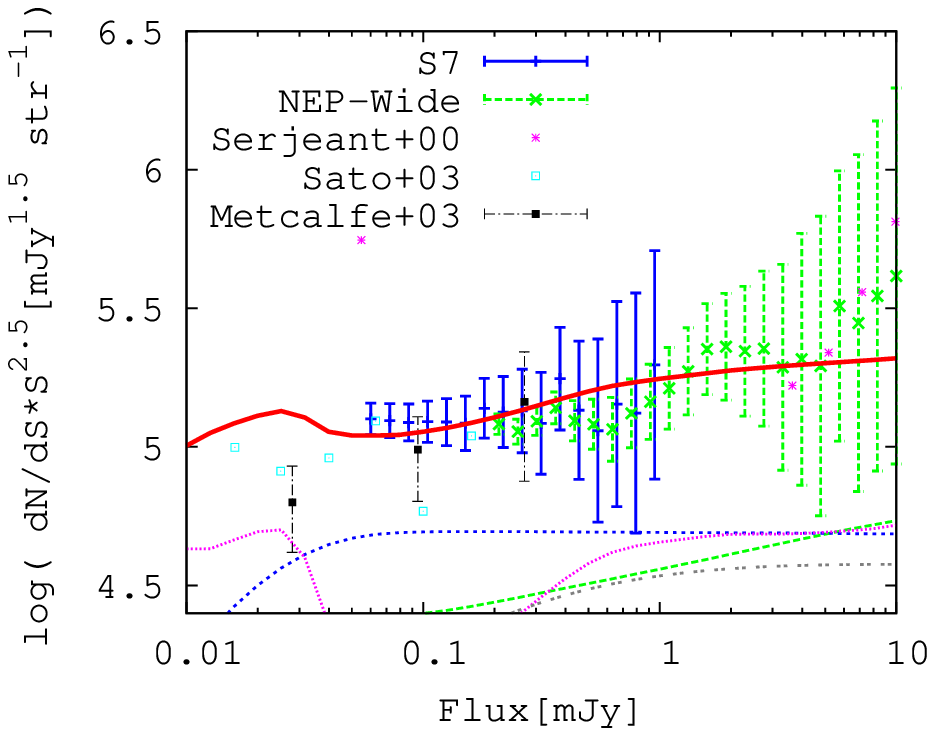}\\
\includegraphics[]{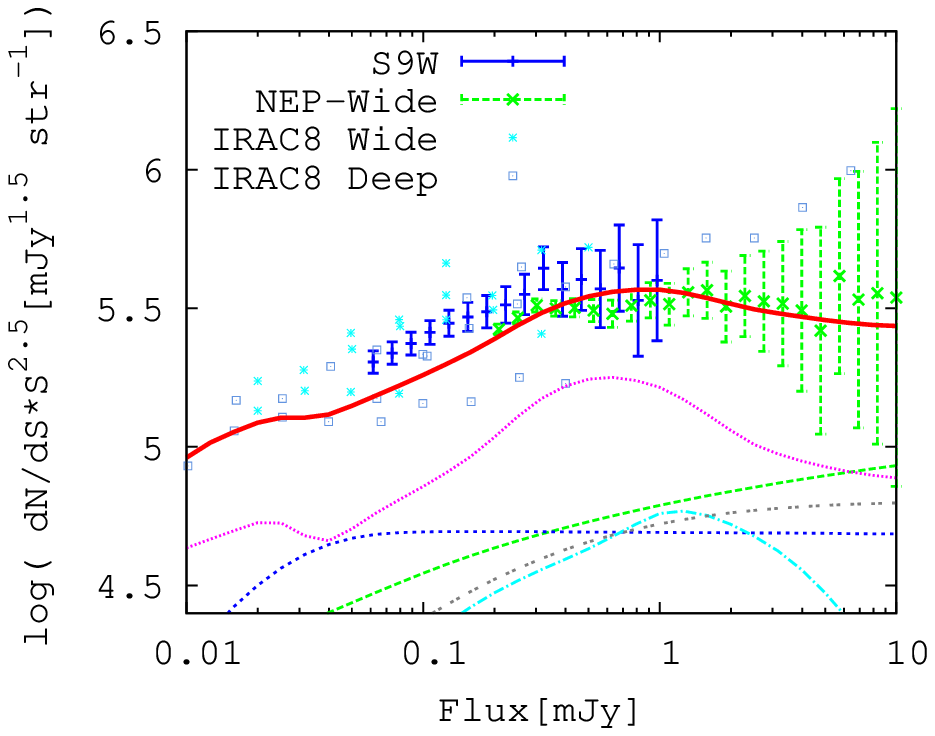}\\
\includegraphics[]{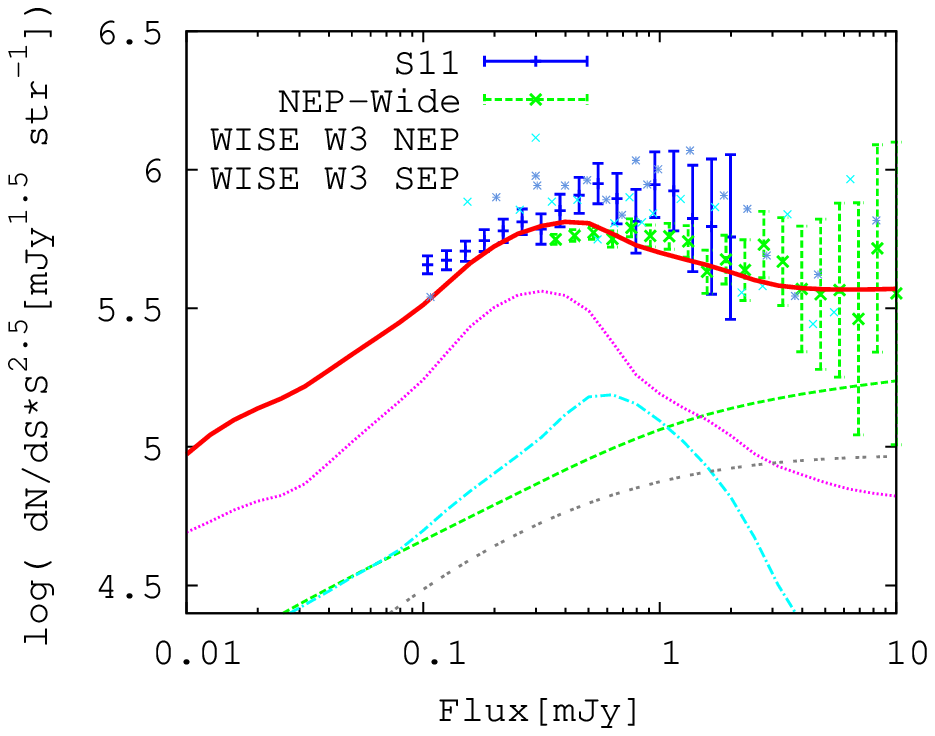}\\
\caption{Differential-source counts from MIR-S bands.
\label{fig:numcnts}
}
\end{figure*}

\begin{figure*}
\centering
\includegraphics[]{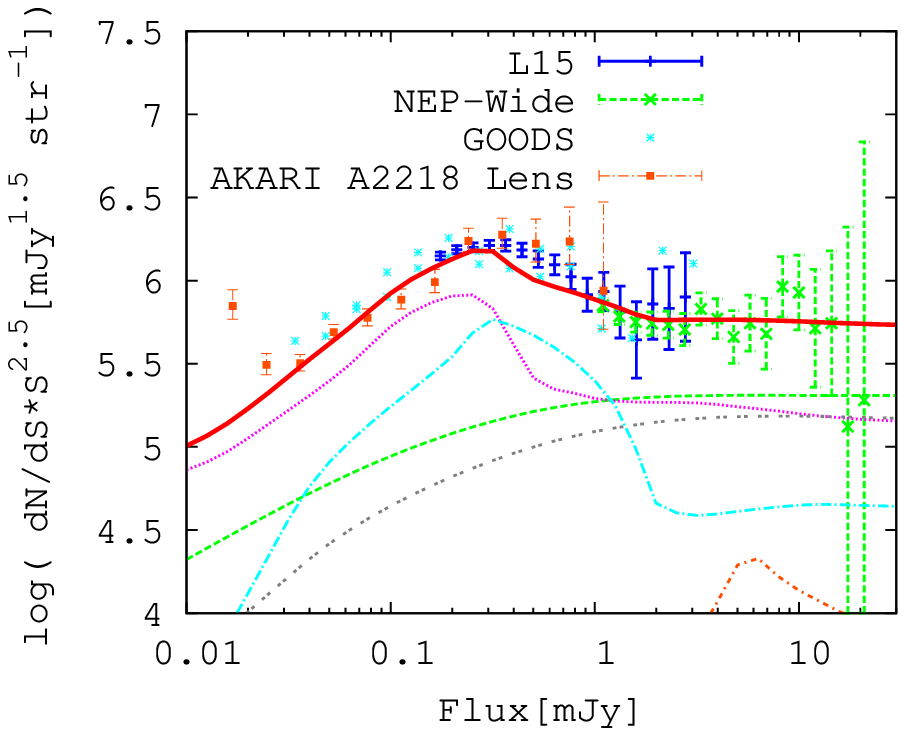}\\
\includegraphics[]{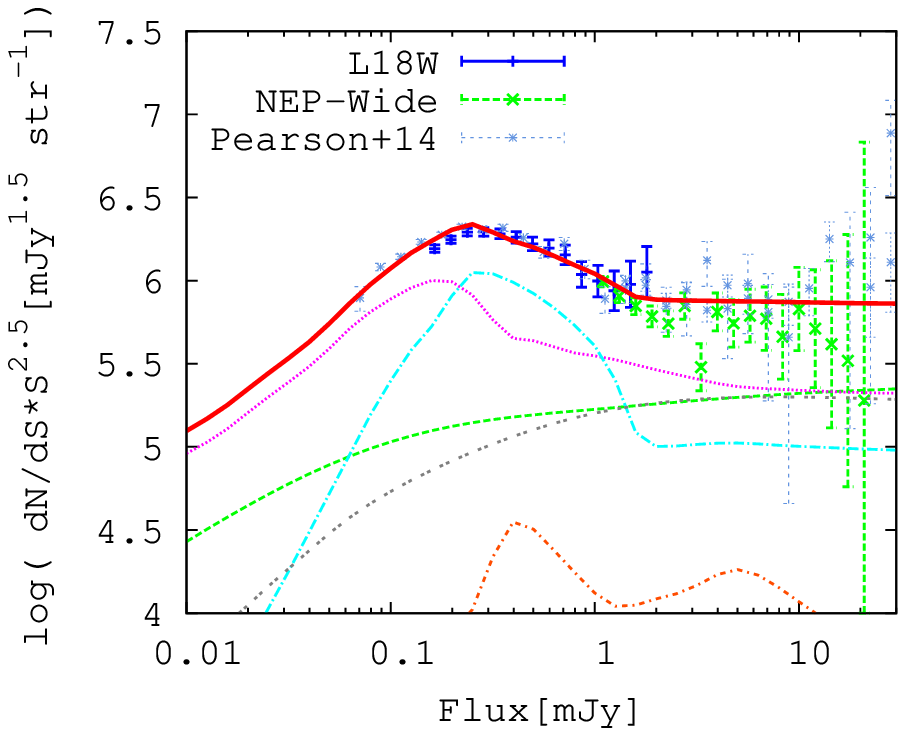}\\
\includegraphics[]{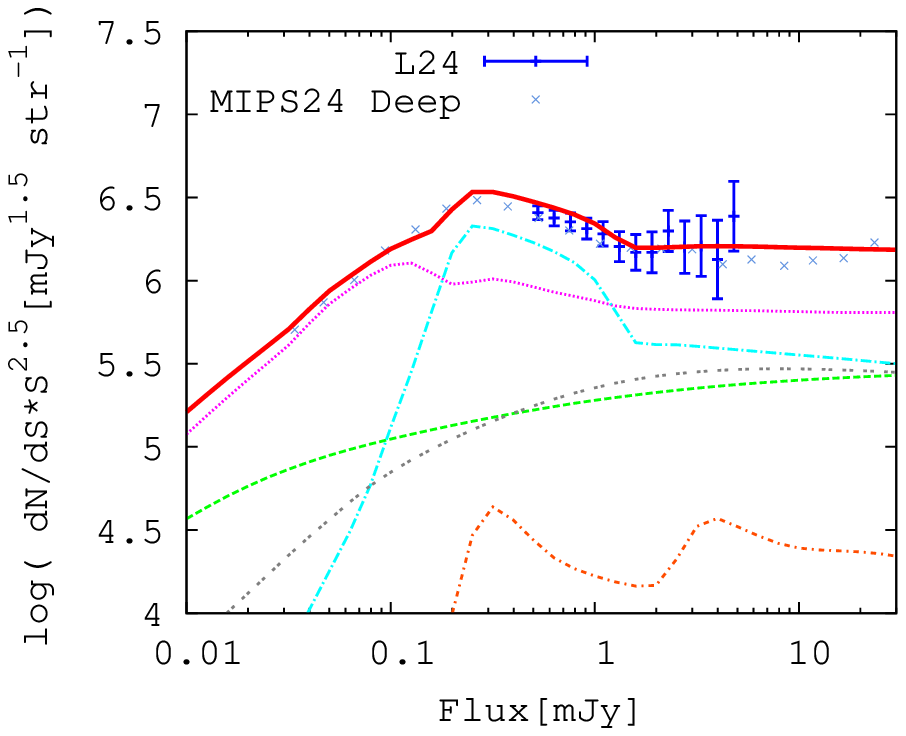}\\
\caption{Differential source counts from MIR-L bands.
\label{fig:numcntl}
}
\end{figure*}

We now provide Euclidean differential source counts of all nine AKARI/IRC bands.
The bright side of the source counts was covered by the NEP-wide survey, and the faint side by the NEP-deep survey.
The completeness correction and the stellar subtraction described in the previous section were applied and the counts were normalised by the areal coverage of the NEP-deep and NEP-wide surveys.
The Euclidean normalised counts are given by $(dN/dS) S^{2.5}$, which highlights any deviation from the Euclidean universe.
Figures \ref{fig:numcntn}, \ref{fig:numcnts} and \ref{fig:numcntl} show the normalised differential source counts of NIR, MIR-S and MIR-L bands for both NEP-deep (blue) and NEP-wide (green).
The NEP-deep and NEP-wide counts are consistent with each other.
\par
For all band counts, the distribution of the bright side is flat (i.e. the slope is 1.5), consistent with an Euclidean universe.
The error bars were calculated with Poisson statistics.
The large error is from stellar subtraction and small-number statistics.
Because we applied the stellar fraction to stellar subtraction, Poisson statistics includes the number of stars and galaxies, which leads to larger errors at brighter flux.
\par
On the fainter side, the behaviour was different in the various bands.
The N2, N3 and N4 counts deviate from the Euclidean slope at $S$$\sim$0.2 mJy.
If the stellar fraction of NIR counts were underestimated (see section \ref{fste}), the galaxy counts would be overestimated.
  However, even if we did not use the criterion of $r'$ $<$ 19, the deviation can be seen; hence the deviation is not due to the uncertainty in the stellar fraction.
These deviations are shown even if we did not correct the completeness. 
Considering the gradual decrease of the completeness for the NIR band, the detection was affected by source confusion even at 0.2 mJy (Fig.\ref{fig:comp}); hence such a deviation is possibly an artificial structure.
The S7 count was almost flat and the S9W count decreased slightly with flux.
The S11 count had a weak hump at $S$$\sim$0.4mJy.
The L15 and L18W counts had a clear hump at $S$$\sim$0.3mJy.
The L24 count deviated from the Euclidean slope, but the detection limit was insufficient to cover the hump.
\par
Notably, the stellar fraction was estimated on the basis of optical data.
The stellarity is known to be useful to determine stars; however, because the NEP-field is located at low galactic latitude $b$$\sim$30$^\circ$, and the stellar fraction is high (Fig. \ref{fig:starfrac}), the unknown uncertainty of the stellar subtraction might affect our results.
Nevertheless, at wavelengths longer than S7-band wavelength, the stellar fraction is low (Fig.\ref{fig:starfrac}); hence, the S9W-L24 counts are quite reliable.
Therefore, the deviation from the Euclidean slope suggests the evolution of the galaxy luminosity function and/or galaxy population.

\subsection{Comparison with other surveys}
Previous surveys that used space- and ground-based telescopes provided galaxy counts in bands similar to the bands used in the present study.
These include the $Ks$ band counts from the WIRCam-deep survey \cite[]{2012A&A...545A..23B},
deep and wide infrared array camera (IRAC) 3.6, 4.5, 5.8, and 8 $\mu$m counts \cite[]{2004ApJS..154...39F}, 
ISO 6.7 $\mu$m counts \cite[]{2000MNRAS.316..768S,2003A&A...405..833S,2003A&A...407..791M}, 
{\it WISE} North/South Ecliptic Pole survey 12$\mu$m count \cite[]{2011ApJ...735..112J},
{\it ISO} 15 $\mu$m count \cite[]{1999A&A...343L..65A,1999A&A...351L..37E,2005MNRAS.358..397V},
IRS 16 $\mu$m survey at GOODS-N/S \cite[]{2011AJ....141....1T}, 
and the deep {\it Spitzer} MIPS 24 $\mu$m counts of \cite{2004ApJS..154...70P}.
Also available are the {\it AKARI} L15, and L18W counts from Hopwood et al. (2010) and Pearson et al. (2010,2014). 
The present work provides the counts of the AKARI/IRC nine bands. 
In this section we compare our counts with the previous counts.
\par
Figure \ref{fig:numcntn} compares the N2, N3 and N4 counts with those of the {\it Ks}, 3.6 and 4.5$\mu$m bands.
For all bands, the bright sides of the counts are consistent with the previous counts.
However, our N2 counts deviate significantly at $S$$\sim$0.2 mJy with respect to the {\it Ks} counts.
Such a deviation can be seen in the N3 and N4 bands as well as in the IRAC bands.
Considering the gradual decrease of the completeness for the NIR band, the detection must have been affected by source confusion even at 0.2 mJy (Fig.\ref{fig:comp}); hence, we cannot reject the possibility that such a deviation is an artificial structure.
\par
The middle panel of Fig. \ref{fig:numcnts} compares the IRAC 8 $\mu$m count with our S9W count.
The two agree very well, despite of having different central wavelengths and band widths.
Both counts decrease monotonically with flux at $S$$<$0.3 mJy.
\par
The bottom panel of Fig. \ref{fig:numcnts} compares the S11 counts with the {\it WISE} 12 $\mu$m counts.
These are generally consistent with each other.
Because our S11 catalogue is approximately two times deeper than that of WISE, a weak hump is observed at $S$$\sim$0.4 mJy.
The top panel of Fig. \ref{fig:numcnts} compares the S7 count with the {\it ISO} 6.7 $\mu$m count; again the two are consistent with each other. 
\par
The top panel of Fig.\ref{fig:numcntl} compares our L15 counts with the {\it Spitzer} IRS 16 $\mu$m and the other {\it AKARI} counts.
These counts agree very well.
The IRS observation concentrated on a small area of GOODS-South (150 arcmin$^2$) and GOODS-North (130 arcmin$^2$); hence, the detection limit is deeper than ours, yet fewer objects were detected than in the present study.
The situation is the same in the {\it AKARI} Abel 2218 field, where the area is small (340 arcmin$^2$), despite the better sensitivity of 0.04 mJy.
Thus, the IRS and {\it AKARI} lens field are appropriate for determining the slope below $S$$\sim$0.1 mJy, whereas the counts from NEP-surveys are useful for determining the hump at $S$$\sim$0.3mJy.
Although we did not show other {\it AKARI} and {\it ISO} counts in this figure (for simplicity), ours are consistent with their counts.
\par
The middle panel of Fig. \ref{fig:numcntl} compares the L18W counts from Pearson et al.(2014) with our counts.
Although they used the same {\it AKARI} NEP-deep data \cite[]{2008PASJ...60S.517W}, the image analyses is different.
Nonetheless, their counts are consistent with ours.
\par
The bottom panel of Fig. \ref{fig:numcntl} compares our L24 count with the MIPS 24 $\mu$m counts.
The L24 count is consistent with the MIPS count.
Since our detection limit is shallower (approximately 0.3 mJy) than that for MIPS count (approximately 0.06 mJy), we cannot cover the slope at fainter flux.

\subsection{Comparison with evolutionary models}
We presented galaxy number counts in nine AKARI 2-24 $\mu$m bands.
Comparing our results with galaxy evolution models is valuable to understand the evolution of the galaxy spectral energy distribution (SED) and population. 
For the galaxy population, we used the spectral libraries of Efstathiou, which provide robust templates from millimetre to near-infrared wavelengths. 
\par
In this work, we adopt the galaxy evolution framework of  \cite{2005MNRAS.358.1417P} and \cite{2009MNRAS.399L..11P}.
The models presented in \cite{2005MNRAS.358.1417P} provided some of the first best fits to the combined MIR source-count observations of {\it ISO} and {\it Spitzer} at 15 $\mu$m and 24 $\mu$m.
For the present work the models were extended to cover a broader range of SEDs by using the spectral libraries of \cite{2003MNRAS.343..322E} and \cite{2000MNRAS.313..734E}.
The model we use includes five populations of galaxies ranging from normal quiescent galaxies to starburst galaxies (defined as $L_{IR}<10^{11}L_{\sun}$) , luminous infrared galaxies (LIRG,$10^{11}L_{\sun}<L_{IR}<10^{12}L_{\sun}$), ultra-luminous infrared galaxies (ULIRG, $L_{IR}>10^{12}L_{\sun}$), and AGN. 
\par
The spectral templates for the quiescent normal galaxy population were selected from the libraries described in \cite{2003MNRAS.343..322E}.
The normal galaxy (often referred to as {\it cirrus}) models of Efstathiou are based on an input interstellar radiation field assuming a stellar spectrum that varies as $\nu S_{\nu} \propto \nu^{3}$ longward of 2.5$\mu$m, plus an assumed interstellar dust field and subsequent radiative-transfer treatment of these two components.
The free parameters of the model are $\chi$, the ratio of radiation field to the local solar neighbourhood and the visual extinction $A_V$.
We select two normal galaxy SEDs from the libraries as templates for the models.
The two selected models assume a $\nu^{3}$ optical input spectrum, with an age of 12.5Gyrs with $A_V$ =1.3 and values of  $\chi$=5 and $\chi$=7 respectively.
\par
Our adopted starburst, LIRG and ULIRG SEDs are taken from the starburst models of  \cite{2000MNRAS.313..734E}.
Their radiative transfer models provide a grid of starburst galaxy SEDs that consider the evolution of an ensemble of optically thick giant molecular clouds (GMCs), illuminated by embedded massive stars. 
The templates assume a Salpeter initial mass function (IMF).
The evolution of the stellar population within the clouds follows the stellar synthesis population models of \cite{1993ApJ...405..538B}.
The individual models are defined by two parameters: the age {\it t\ } of the starburst in Myr, and the initial optical depth $\tau$ (in {\it V\ }) of the molecular clouds. 
These models have been shown to provide good fits to selected samples of IRAS, ISO and Spitzer starbursts and ultra-luminous galaxy populations \cite[]{2000MNRAS.319.1169E,2004MNRAS.351.1290R,2005AJ....129.1183R}.
In general, a higher optical depth implies a higher far-infrared luminosity (i.e. more of the hot ultra-violet starlight is absorbed by the dust). 
We selected a total of seven SEDs (two starbursts, three LIRGs, two ULIRGs) from the template libraries to ensure a reasonable variation in the the mid-infrared PAH features in an attempt to avoid features in the source counts caused by a particular SED choice for all sources. 
\par
For our AGN template, we use the tapered disc dust torus model of \cite{1995MNRAS.273..649E}, with the NIR-optical spectrum following \cite{2003MNRAS.339..260K}.
We also include an elliptical component for the model with an SED taken from \cite{2000A&A...363..517M}.
\par
To model the number density per luminosity bin of our model populations, we use the 60 $\mu$m type-dependent luminosity function of \cite{1990MNRAS.242..318S} derived from the $IRAS$ surveys described in \cite{2000MNRAS.317...55S} for the normal and star-forming (starburst, LIRG and ULIRG ) populations.
This is effectively a zero-redshift local luminosity function, which is then evolved with redshift by using parametric functions.
The AGN and elliptical populations are modelled separately by using the luminosity functions of \cite{2006A&A...451..443M} and Table 3 of \cite{2001ApJ...560..566K}, respectively.
\par
The evolutionary scenario for our galaxy populations assumes a rapid onset of star-formation at high redshift, a gradual decline to a redshift of approximately unity, and then a rapid decline in star-formation activity to the present epoch (e.g. Elbaz 2005, Le Floc'h et al. 2005, Goto et al. 2010, Gruppioni et al. 2013, Patel et al. 2013).
Notably, our population classification differs somewhat from that of \cite{2013MNRAS.432...23G}, who modelled the far-infrared population at longer wavelengths ($>$70$\mu$m); however, both models attribute strong luminosity evolution and positive density evolution to the starburst and LIRG-ULIRG populations. 
The relative contribution of each population component to the overall star-formation rate as a function of redshift follows the overall downsizing pattern, in which the most-massive galaxies formed stars at an early epoch, thus dominating the star-formation history of the Universe at high redshift (e.g. Mobasher et al. 2009).
The model assumes that all active populations evolve steeply in both luminosity, $F(z)$, and density, $G(z)$, as a double power law to a redshift of unity and are somewhat shallower thereafter to z$\sim$3 as $F(z), G(z) =(1+z)^{f,g}$.
Thus, four variables are required: power index from redshift zero to the first evolutionary peak at z$\sim$1 and power index to a higher redshift z$\sim$3, a pair for each of both the luminosity evolution ($F(z)$ ($f_{1}$, $f_{2}$) and the density evolution $G(z)$ ($g_{1}$, $g_{2}$)).
For redshifts z $>$3, the evolutionary power is assumed to decrease exponentially.
The adopted parameters for each galaxy population are summarised in Table \ref{evolution}.

\begin{table}
\caption{Galaxy-evolution parameters for the models described in the text.
  The extragalactic population is modelled on six generic galaxy types.
  For population, there are two sets of evolutionary parameters corresponding to luminosity evolution $f_{1}, f_{2}$ and density evolution $g_{1}, g_{2}$ to redshift z$_{1}$=1 and z$_{2}$=3, respectively.}
\begin{tabular}{@{}lcccc}
\hline\noalign{\smallskip}
Component	&	$f_{1}$	& $f_{2}$	&	$g_{1}$	&	$g_{2}$	\\
\hline
Normal    	&	0	&	0	&	0	&	0	\\
Elliptical   	&	0	&	0	&	0	&	0	\\
Starburst 	&	3.2	&	2.2	&	1	&	0.4	\\
LIRG       	&	2.5	&	2.2	&	3.2	&	2.5	\\
ULIRG     	&	2.5	&	2.2	&	3.2	&	2.5	\\
AGN      	&	2	&	2.2	&	0	&	0	\\
\hline
\end{tabular}\\
\label{evolution}
\end{table}
\smallskip

\par
In Figures \ref{fig:numcntn}, \ref{fig:numcnts} and \ref{fig:numcntl}, the evolutionary model (total counts and individual components) is compared with the {\it AKARI} source counts presented in this work along with ancillary data from {\it ISO}, {\it Spitzer} and ground-based observations.
The near-infrared counts in Figure \ref{fig:numcntn} are completely dominated by the non-evolving normal spiral and elliptical populations.
The evolutionary model fits both the Euclidean regime sampled by the {\it AKARI} source counts and the fainter turnover probed by the deeper ground-based $K$-band and {\it Spitzer}-IRAC surveys.
The IRC MIR-S S7, S9W and S11 bands shown in Figure \ref{fig:numcnts} track the emergence of the strongly evolving star-forming population responsible for the evolutionary bump in the source counts prominent in the S9W and S11 bands.
In the S11 band, from the non-evolving populations, we see a negligible contribution to the counts that is fainter than a few mJy. 
At longer mid-infrared wavelengths, the source counts are especially well sampled by {\it ISO}, {\it Spitzer} and {\it AKARI} at 15 and 24 $\mu$m.
The evolutionary model satisfactorily reproduces the rise in the source counts at the $\sim$mJy level, the peak in the counts between 0.1 and 1 mJy and the decline in the counts at fainter flux levels.
The model predicts that the emerging LIRG population become dominant from 15 to 24 $\mu$m.
There is also a non-negligible contribution from AGN at fluxes greater than approximately 1 mJy.
\par
By integrating the models, we calculated the total cosmic infrared background intensity in each band.
The resulting intensities are 8.9, 6.0, 4.0, 1.9, 1.7, 1.9, 2.3, 2.6 and 3.0 nW m$^{-2}$ sr$^{-1}$ for the bands N2, N3 N4, S7, S9W, S11, L15, L18W and L24, respectively.
When we integrate the model down to 80\% completeness for each band, we obtain 3.2, 1.5, 0.91, 0.55, 0.71, 0.85, 1.0, 1.2 and 0.75 nW m$^{-2}$ sr$^{-1}$, respectively.
This result implies that the AKARI NEP-deep survey has resolved 36\%, 26\%, 22\%, 29\%, 41\%, 44\%, 44\%, 45\% and 25\% of the cosmic infrared background at 2.4, 3.2, 4.1, 7, 9, 11, 15, 18 and 24 $\mu$m.

\section{Summary}
We present galaxy counts for observations in all nine AKARI/IRC bands of the NEP-deep and NEP-wide surveys.
The counts from the NEP-deep survey are presented down to the 80\% completeness limits; 0.18, 0.16, 0.10, 0.05, 0.06, 0.10, 0.15, 0.16, and 0.44 mJy in the N2, N3, N4, S7, S9W, S11, L15, L18W, and L24 bands, respectively. 
The counts on the brighter side are covered by the NEP-wide survey.
The completeness and the difference between observed and intrinsic magnitudes are corrected by using a Monte Carlo simulation and a $P_{ij}$ matrix.
The stellar counts are subtracted with the stellar fraction estimated by using the optical catalogue for a 2 deg$^2$ area of the NEP-wide field. 
The N2, N3 and N4 counts are consistent with $Ks$, 3.6 and 4.5 $\mu$m counts on the bright side although the faint side disagrees probably because of source confusion.
The distribution of S7 counts is flat and is consistent with the ISO counts.
The S9W counts slightly decrease with flux and are consistent with IRAC 8$\mu$m band counts.
The deep detection limit of the S11 band allows us to observe a weak hump at $S$$\sim$0.4mJy.
The L15 and L18W counts also have a hump at $S$$\sim$0.3mJy, consistent with previous studies.
The L24 counts partially cover the hump at $S$$\sim$0.3mJy that was found in the previous study.
Galaxy evolutionary models satisfactorily reproduce these number counts.
By integrating the models, the cosmic infrared background intensity in the nine bands is calculated, based on which we conclude the AKARI NEP-deep survey resolves 20\%-50\% of the cosmic infrared background.

\section*{Acknowledgements}
The {\it AKARI} NEP-Deep survey project activities are supported by JSPS grant 23244040. 
The authors would like to thank Enago (www.enago.jp) for the English language review.
\bibliographystyle{mn2e}
\bibliography{../BIB/scount,../BIB/ref,../BIB/fagn}

\bsp

\label{lastpage}

\end{document}